\documentclass[sigconf]{acmart}

\usepackage{subcaption, balance}

\AtBeginDocument{%
  \providecommand\BibTeX{{%
    \normalfont B\kern-0.5em{\scshape i\kern-0.25em b}\kern-0.8em\TeX}}}

\copyrightyear{2021}
\acmYear{2021}
\setcopyright{acmcopyright}\acmConference[KDD '21]{Proceedings of the 27th ACM SIGKDD Conference on Knowledge Discovery and Data Mining}{August 14--18, 2021}{Virtual Event, Singapore}
\acmBooktitle{Proceedings of the 27th ACM SIGKDD Conference on Knowledge Discovery and Data Mining (KDD '21), August 14--18, 2021, Virtual Event, Singapore} \acmPrice{15.00}
\acmDOI{10.1145/3447548.3467175}
\acmISBN{978-1-4503-8332-5/21/08}

\begin{document}
\fancyhead{}

\title{We Know What You Want: An Advertising Strategy Recommender System for Online Advertising}
\author{Liyi Guo$^1$, Junqi Jin$^2$, Haoqi Zhang$^1$, Zhenzhe Zheng*$^{1,3}$, Zhiye Yang$^2$, Zhizhuang Xing$^2$, Fei Pan$^2$
\and
Lvyin Niu$^2$, Fan Wu$^1$, Haiyang Xu$^2$, Chuan Yu$^2$, Yuning Jiang$^2$, Xiaoqiang Zhu$^2$}
\affiliation{
\institution{$^1$Shanghai Jiao Tong University, $^2$Alibaba Group\\$^3$State Key Lab. for Novel Software Technology, Nanjing University}
\and
\{liyiguo1995, zhanghaoqi39, zhengzhenzhe\}@sjtu.edu.cn, fwu@cs.sjtu.edu.cn
\and
\{junqi.jjq, zhiye.yzy, zhizhuang.xzz, pf88537, lvyin.nly, shenzhou.xhy, yuchuan.yc, mengzhu.jyn, xiaoqiang.zxq\}@alibaba-inc.com
\country{}
}

\thanks{This work was supported in part by Science and Technology Innovation 2030 – ``New Generation Artificial Intelligence'' Major Project No. 2018AAA0100905, in part by China NSF grant No. 62025204, 62072303, 61902248, and 61972254, and in part by Alibaba Group through Alibaba Innovation Research Program, and in part by Shanghai Science and Technology fund 20PJ1407900. The opinions, findings, conclusions, and recommendations expressed in this paper are those of the authors and do not necessarily reflect the views of the funding agencies or the government.}
\thanks{*Z. Zheng is the corresponding author.}

\renewcommand{\shortauthors}{Guo et al.}

\begin{abstract}
Advertising expenditures have become the major source of revenue for e-commerce platforms. Providing good advertising experiences for advertisers by reducing their costs of trial and error in discovering the optimal advertising strategies is crucial for the long-term prosperity of online advertising. To achieve this goal, the advertising platform needs to identify the advertiser's optimization objectives, and then recommend the corresponding strategies to fulfill the objectives. In this work, we first deploy a prototype of strategy recommender system on Taobao display advertising platform, which indeed increases the advertisers' performance and the platform's revenue, indicating the effectiveness of strategy recommendation for online advertising. We further augment this prototype system by explicitly learning the advertisers' preferences over various advertising performance indicators and then optimization objectives through their adoptions of different recommending advertising strategies. We use contextual bandit algorithms to efficiently learn the advertisers' preferences and maximize the recommendation adoption, simultaneously. Simulation experiments based on Taobao online bidding data show that the designed algorithms can effectively optimize the strategy adoption rate of advertisers.

\end{abstract}



\maketitle

\begin{small}
\textbf{ACM Reference Format:}\\
Liyi Guo, Junqi Jin, Haoqi Zhang, Zhenzhe Zheng, Zhiye Yang, Zhizhuang Xing, Fei Pan, Lvyin Niu, Fan Wu, Haiyang Xu, Chuan Yu, Yuning Jiang, Xiaoqiang Zhu. 2021. We Know What You Want: An Advertising Strategy Recommender System for Online Advertising.
In \emph{Proceedings of the 27th ACM SIGKDD Conference on Knowledge Discovery and Data Mining (KDD '21), August 14--18, 2021, Virtual Event, Singapore}. ACM, New York, NY, USA, 9 pages. https://doi.org/10.1145/3447548.3467175
\end{small}

\pagestyle{plain} 

\section{Introduction}
With the rapid development of the Internet, online e-commerce platforms, such as Amazon~\cite{linden2003amazon}, eBay~\cite{resnick2006value}, and Taobao~\cite{zhu2017optimized}, have become the major venues for people to buy and sell products. E-commerce advertising is a critical marketing tool for advertisers to efficiently deliver their products to potential buyers, and the revenue from advertising has become the major source of income for e-commerce platforms~\cite{lahaie2007sponsored, evans2008economics, evans2009online, goldfarb2011online}. Thus, providing good advertising experiences for advertisers is crucial to the long-term prosperity of online advertising. It is highly necessary for platforms to fully understand the optimization objectives of advertisers, and launch advertising services to fulfill them.

\begin{figure}[t]
  \centering
  \includegraphics[width=1.0\linewidth]{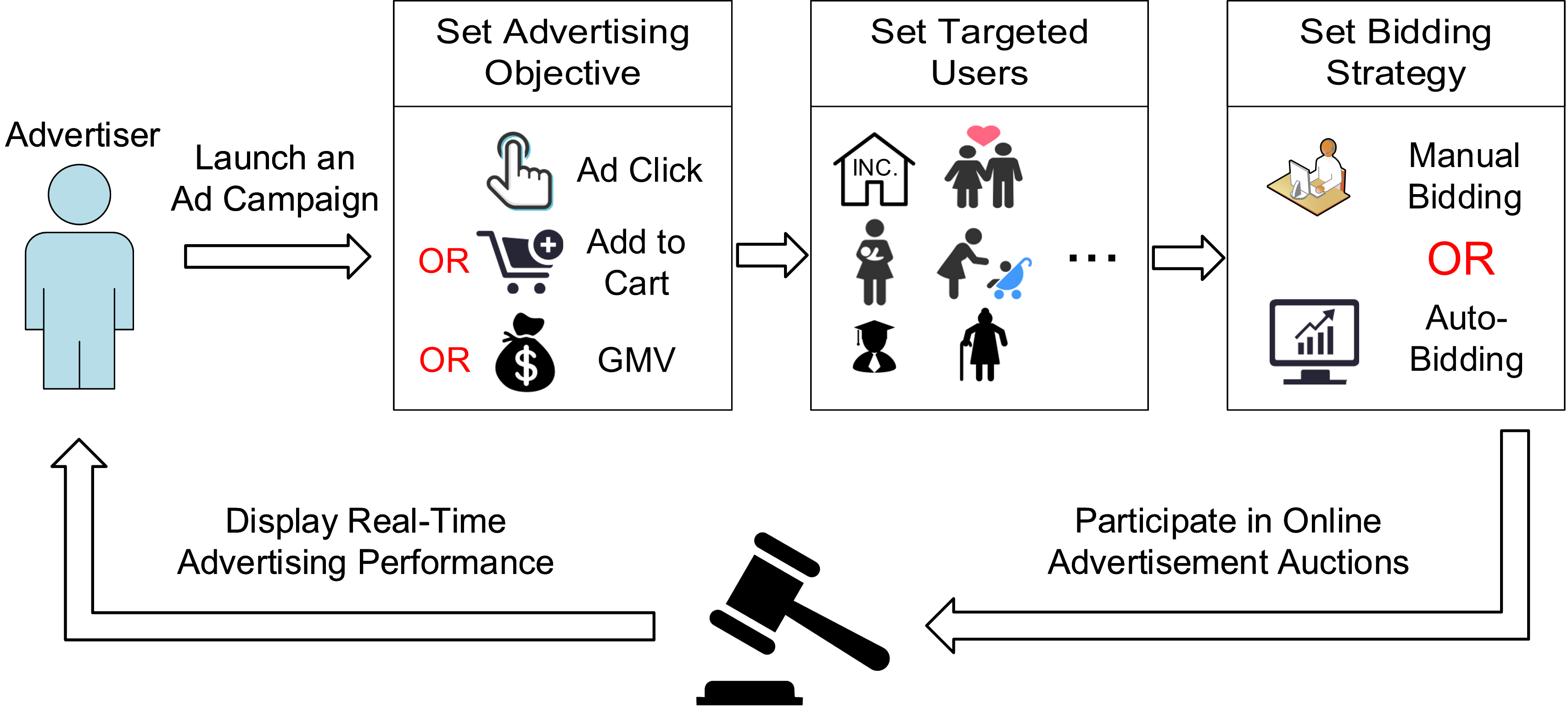}
  \caption
    {The advertising process in Taobao display advertising platform.}
\label{advertising_process}
\end{figure}

As a representative e-commerce advertising platform, Taobao has established an advanced online advertising system with various intelligent services to decide the display of tens of millions of online ads. For the advertisers' side, in addition to traditional manual bidding, the Taobao display advertising system also provides auto-bidding services, such as Budget Constraint Bidding (BCB) and Multiple Constraint Bidding (MCB), to help advertisers fulfill their advertising objectives in a cost-efficient way. Since the performance achieved by the intelligent advertising strategies highly depends on the preset optimization objectives, the platform would like to first identify the advertisers' preference for various advertising performance. The current launching process of an ad campaign shown in Figure~\ref{advertising_process} contains three steps: (1) Choosing an optimization objective, such as the number of clicks, gross merchandise volume (GMV), etc. (2) Selecting the targeted users. (3) Choosing manual bidding or auto-bidding services. However, such a self-service launching process would introduce some obstacles due to the information asymmetry between advertisers and the platform. On the one hand, the advertisers have no access to the real-time information of online ad auctions, which makes them difficult to set appropriate advertising strategies, leading to poor performance and the loss of budget. On the other hand, the current number of advertising objectives for advertisers to choose is limited, which does not capture all possible optimization objectives of advertisers. The current advertising system does not have channels for advertisers to reveal fine-grained optimization objectives. A recent survey reveals that a large number of new advertisers have left the platform because their advertising objectives were not well satisfied~\cite{guo2020deep}. To solve the above problems and further improve the performance of the advertising platform, we initialize a new research direction: designing advertising strategy recommender systems for advertisers.

An intuitive approach to implement an advertiser-side recommender system is to borrow ideas from the extensive works in user-side recommender system, which focuses on the matching between products and users~\cite{cheng2016wide, covington2016youtube, guo2017deepfm, zhou2018deep, zhu2018learning}. However, simply adopting these approaches to the advertiser-side recommendation would encounter several new challenges: (1) User-side recommendation applies standard supervised learning methods based on users' click or purchase behaviors, which are the direct and explicit signals for users to express their interests and preferences. While in the current advertising platforms, due to the lack of efficient and accurate estimation for advertising performance, the platforms do not provide the predicted performance of recommended strategies, and thus advertisers cannot express their preferences for advertising performance. (2) Most models on the user-side recommendation are based on a discrete corpus, that is the product library. However, on the advertiser-side, the corpus is often seen as a complex high-dimensional continuous space, which represents the potential performance results from different advertising strategies. (3) There are no available public data set yet for training the model of recommender systems for advertisers. It is also expensive for the recommender system to explore an advertiser's advertising objective, because the advertiser's adoption of new strategies would expend her advertising cost immediately.

To overcome the above challenges, we first design a prototype recommender system to help advertisers optimize their bid prices and select the targeted users. Online evaluations demonstrate that the prototype system can indeed increase the advertiser's advertising performance and the platform's revenue. We further augment this prototype recommendation system by the following three extensions. To solve the difficulty of learning advertisers' optimization objectives, we recommend the advertising strategies associated with the predicted performance, which is obtained through the ad auction simulator. We model the advertiser's preference as the weight for each individual performance indicator, and the resulting optimization objective is the linear combination of performance indicators. We learn the weights through the advertisers' interactions with the recommended strategies. For the second challenge, we design a neural network with the input of advertisers' features and their historical adapted strategies, to predict the advertisers' potential preferences, reducing the complexity of searching over continuous space. For the last challenge, to effectively learn advertisers’ preferences and objectives at the same time, we consider the recommended strategies as bandits and the advertiser’s strategy adoption as the environmental feedback, and leverage the techniques from contextual bandit to solve this problem.

We summarize the contributions of this work as follows:

$\bullet$ To the best of our knowledge, we are the first to investigate the problem of advertising strategy recommendation for online advertising.
We define the objective of an advertiser is to maximize a linear combination of multiple performance indicators, where the weight for each performance indicator is considered as the advertiser's preference.  We recommend the advertising strategies associated with the predicted performance, and use advertisers’ adoption behaviors as signals to infer their preferences.

$\bullet$ To efficiently recommend advertising strategies, we should exploit the existing information to optimize the advertiser's adoption rate and properly explore the advertiser's underlying preference at the same time. We model this explore-and-exploit process as a contextual bandit problem: we regard the recommending advertising strategies as the actions and the advertisers' adoption behaviors as the rewards. We also use the Dropout technique to make a efficient trade-off between exploration and exploitation. 

$\bullet$ We deploy a prototype system recommending bid prices and targeted users to advertisers on the Taobao display advertising platform. The online A/B test demonstrates the effectiveness of recommending advertising strategies for advertisers, increasing the advertiser's  performance and the platform's revenue. We build an offline simulation environment based on Taobao online advertising bidding data, and conduct a series of experiments. The experiment results show that the strategy recommender system we design can effectively learn the advertisers' preferences and improve the recommendation adoption rate of advertisers.

The rest of the paper is organized as follows: In Section 2, we introduce related works about user-side recommender systems and real-time bidding algorithms. In Section 3, we introduce the prototype recommender system deployed in Taobao's advertising platform and the new strategy recommender system designed based on the prototype system. In Section 4, the learning algorithm of the new strategy recommender system is illustrated. The experimental results are given in Section 5. We conclude the work in 
Section 6. 

\section{Related Work}

\subsection{Recommender System}
In recent years, the recommendation problem has been extensively studied in both academia and industry, and different types of recommender systems have been deployed in various scenarios~\cite{cheng2016wide, covington2016youtube, guo2017deepfm, zhou2018deep, zhu2018learning}. The recommender systems resolve the problem of information overload: help users discover useful information quickly and provide personalized services~\cite{lu2015recommender}. Several approaches, such as collaborative filtering~\cite{sarwar2001item, schafer2007collaborative}, content-based filtering~\cite{pazzani2007content, lops2011content} and hybrid filtering~\cite{basilico2004unifying}, have been widely used in designing recommender systems. Collaborative filtering leverages the correlation between products to do recommendation. Content-based filtering recommends products to users based on their similarity, which is extracted from the user's historical choices, user profiles and product information. Hybrid filtering combines collaborative filtering and content-based filtering to improve the accuracy of recommendations.

Although extensive studies have been done on the user-product recommendation, there are two main differences between the user-side recommendation and the advertiser-side recommendation. First, user-product recommendations are usually based on a unified product library for feature extraction and correlation analysis. However, in the advertiser-side recommendation, the recommended product is an abstract advertising strategy (such as bidding strategy), which is hard to analyze simply through feature extraction. Second, advertiser-side recommendations should not only solve the problem of information overload but also help advertisers optimize their advertising performance, achieving the promised performance of the recommended strategies. 
Therefore, we need to jointly consider the advertising strategy recommendation and advertising objectives optimization, which is related to real-time bidding.

\subsection{Real-Time Bidding}

In the real-time bidding scenario, researchers have conducted an in-depth study on the problem of auto-bidding to optimize certain optimization objectives~\cite{zhang2012joint,ren2017bidding,maehara2018optimal}. Designing a real-time bidding algorithm is formulated as an online optimization problem with an optimization goal, e.g. maximize GMV under the budget constraint~\cite{zhang2014optimal,wu2018budget}. 
Zhang et al. proposed an offline optimal bidding strategy by bidding $\frac{v}{\lambda}$ for each impression under the second-price auction mechanism, where $v$ is the value of the impression, and $\lambda$ is a fixed parameter determined by the environment~\cite{zhang2016optimal}. Wu et al. further applied a model-free reinforcement learning method to adjust the $\lambda$ in an online manner, aiming to pace the spend of the budget within the real-time environment~\cite{wu2018budget}. The subsequent research efforts attempted to extend the single objective and single constraint to the setting with various optimization goals under multiple constraints~\cite{zhang2016feedback,kitts2017ad,yang2019bid}. Zhang et al. used feedback control methods to adjust bidding based on real-time performance, and satisfied auction winning rate and pay per click (PPC) constraints~\cite{zhang2016feedback}. Yang et al. formulated the GMV optimization problem with budget and PPC constraints as an online linear programming problem~\cite{yang2019bid}.

Based on the extensive work about real-time bidding, the auto-bidding tools provided by the platform can well handle the optimization problem as long as the objectives are known in advance. However, in a display advertising system, advertisers often have a variety of specific objectives for different ads at different marketing phases, leading to different optimization goals. 
For example, advertisers would like to maximize the amount of impressions for new products, and maximize the GMV of the product in promotional sessions.
However, there still does not exist effective ways for the advertising platforms to figure out the advertiser's advertising objectives. This significantly degrades the performance of the real-time bidding algorithms.

\begin{figure}[t]
  \centering
  \includegraphics[width=1.0\linewidth]{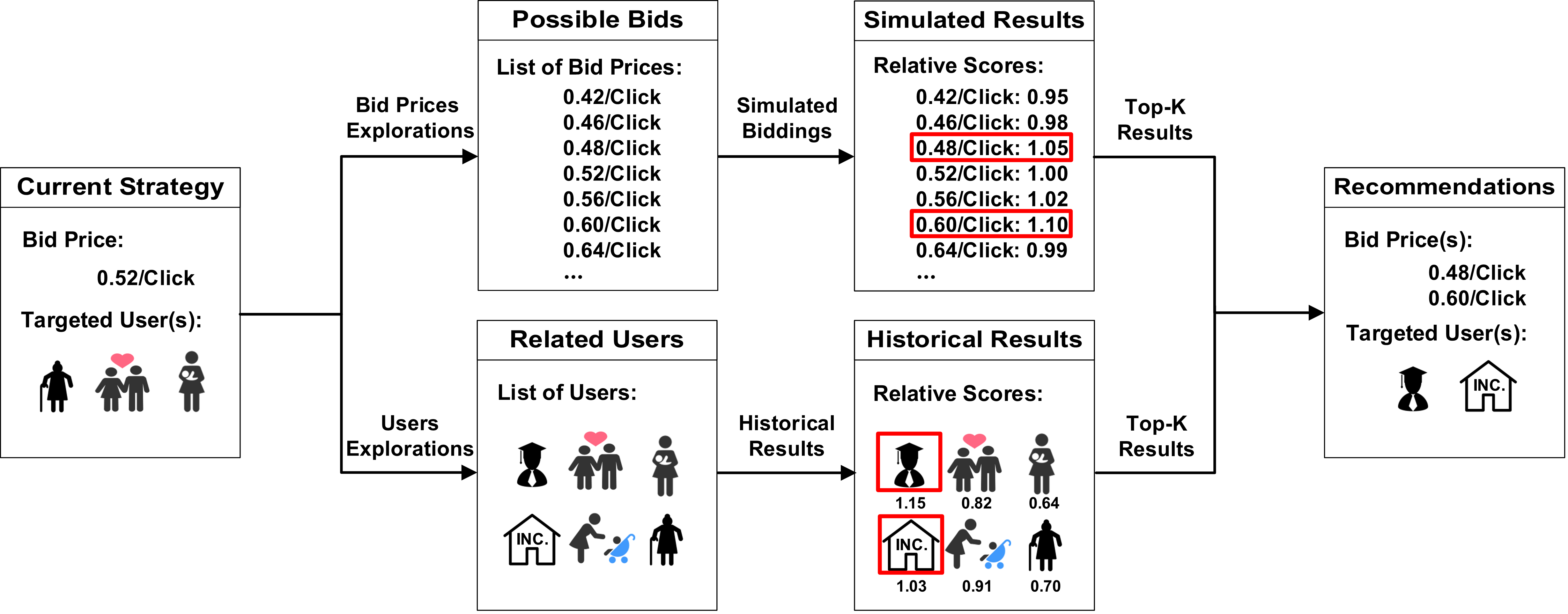}
  \caption
    {The prototype recommender system: a bid optimization module and a targeted user optimization module.}
\label{fig:prototype}
\end{figure}

\section{System Design}
\label{question_modeling}
In this section, we first illustrate our currently deployed advertising strategies recommender system for advertisers, and formally define the advertisers' preference and optimization goals. We then introduce our extensions for this prototype system. Finally, we formulate the problem of advertising strategy recommendation as the contextual bandit, and provide efficient solutions to solve it.

\subsection{Prototype Recommender System}
\label{sec:current_system}
To reduce the expensive trial and error costs for advertisers to look for optimal advertising strategies, and explore potential marketing opportunities, we have already deployed a recommender system on the Taobao display advertising platform as shown in Figure~\ref{fig:prototype}. The system consists of two components: a \emph{bid optimization module} and a \emph{targeted user optimization module}. The bid optimization module recommends bids for advertiser based on her selected targeted users and the pre-defined optimization goal, and the targeted user optimization module explores potential users for the advertiser. The prediction ability of the system is supported by the \emph{ad auction simulator}, which estimates the advertising performance of possible strategies using the advertiser's historical bidding logs through conducting simulated auctions.

\noindent\textbf{Bid Optimization Module.} 
To search for high-quality bids for a certain type of targeted users, the bid optimization module first samples some candidate bids, predicts the advertising performance of these bids through the ad auction simulator. The module ranks these candidate bids by some rule, such as the relative increment on CTR (or CVR) compared with the average CTR (or CVR) of the historical adopted bidding strategies.
The module lists the top-K bidding strategies associated with some other information, such as historical advertising performance, current bids and predicted advertising performance.

\noindent\textbf{Targeted User Optimization Module.} This module explores potential users for advertisers. It first merges similar types of users in a proper way, such as the users selected by advertisers belong to the same product category.
The module calculates the average historical KPIs of these merged users, ranks them  based on some performance indicators (e.g. $CTR$, $PPC$, $CVR$, $GMV$). For each ad campaign, the top-K targeted users would be recommended, and the predicted number of unique visitors for the ad campaign covered by these targeted users would also be shown.

The prototype recommender system increases the ad platform's Average Revenue per User (ARPU) by $1.2\%$ in the online A/B test in May 2020. The implementations of the system, experimental settings, and results of the online evaluations are shown in Section~\ref{systen_online_evaluation}. The currently deployed system is still in the infant stage: (1) The system ranks the candidate advertising strategies based on some deterministic metrics (i.e. relative increment in some KPIs or a fixed weighted average over KPIs). 
However, advertisers have various advertising preferences and objectives over the advertising performance, and ranking recommended strategies based on this information would improve the adoption rate and provide personalized advertising experiences for advertisers. 
(2) The current system gives recommendations based on a fixed bid price. 
We observe that the advertisers care more about the performance of the advertising strategy, instead of the specific contents of the strategy, such as the bid for each auction. We argue that the strategy recommender system for advertisers should focus on the advertising performance.  

\subsection{Advertising Strategy Recommender System}

To further augment the current prototype system, we first formulate the problem of strategy recommendation, and then we design a new recommender system based on the formulation. We define the performance of an ad campaign as: $\mathbf{v} = [v_1, v_2, \cdots, v_n]^\top$, where $n$ represents the number of KPIs, and $v_i$ represents the value of the $i$-th KPI. Since advertisers have different preferences over different KPIs, we define the advertiser's preference as a weight vector: $\mathbf{w} = [w_1, w_2, \cdots, w_n]^\top$, where $w_i$ is the weight for the $i$-th KPI. For a specific ad campaign, suppose that the advertiser's preference is $\mathbf{w}^{*}$, and the corresponding optimization objective of the advertiser is ${\mathbf{w}^{*}}^\top \cdot \mathbf{v}$, which is a weighted sum of advertising performance. By defining $\Pi$ as the advertiser’s bidding strategy, for the advertiser with the preference vector $\mathbf{w}^{*}$, recommending the optimal bidding strategy $\Pi_{\mathbf{w}^{*}}$ is to solve the following optimization problem:
\begin{equation}
\label{optimal_bidding}
    \Pi_{\mathbf{w}^{*}} = \mathop{\arg\max}_{\Pi} {\mathbf{w}^{*}}^\top \cdot \mathbf{v}_{\Pi},
\end{equation}
where $\mathbf{v}_{\Pi}$ is the advertising performance that the bidding strategy $\Pi$ can achieve.
To solve optimization problem in~(\ref{optimal_bidding}), an intuitive solution is to use some real-time bidding algorithms~\cite{zhang2014optimal,zhang2016feedback,cai2017real,wu2018budget} with the preference vector $\mathbf{w}^{*}$. Therefore, the preference vector $\mathbf{w}^{*}$ guides the advertising platform to obtain the most satisfactory advertising strategies and performance for advertisers. However, due to the information asymmetry between the platform and advertisers, we cannot directly know the vector $\mathbf{w}^{*}$.
\begin{figure}[t]
  \centering
  \includegraphics[width=1.0\linewidth]{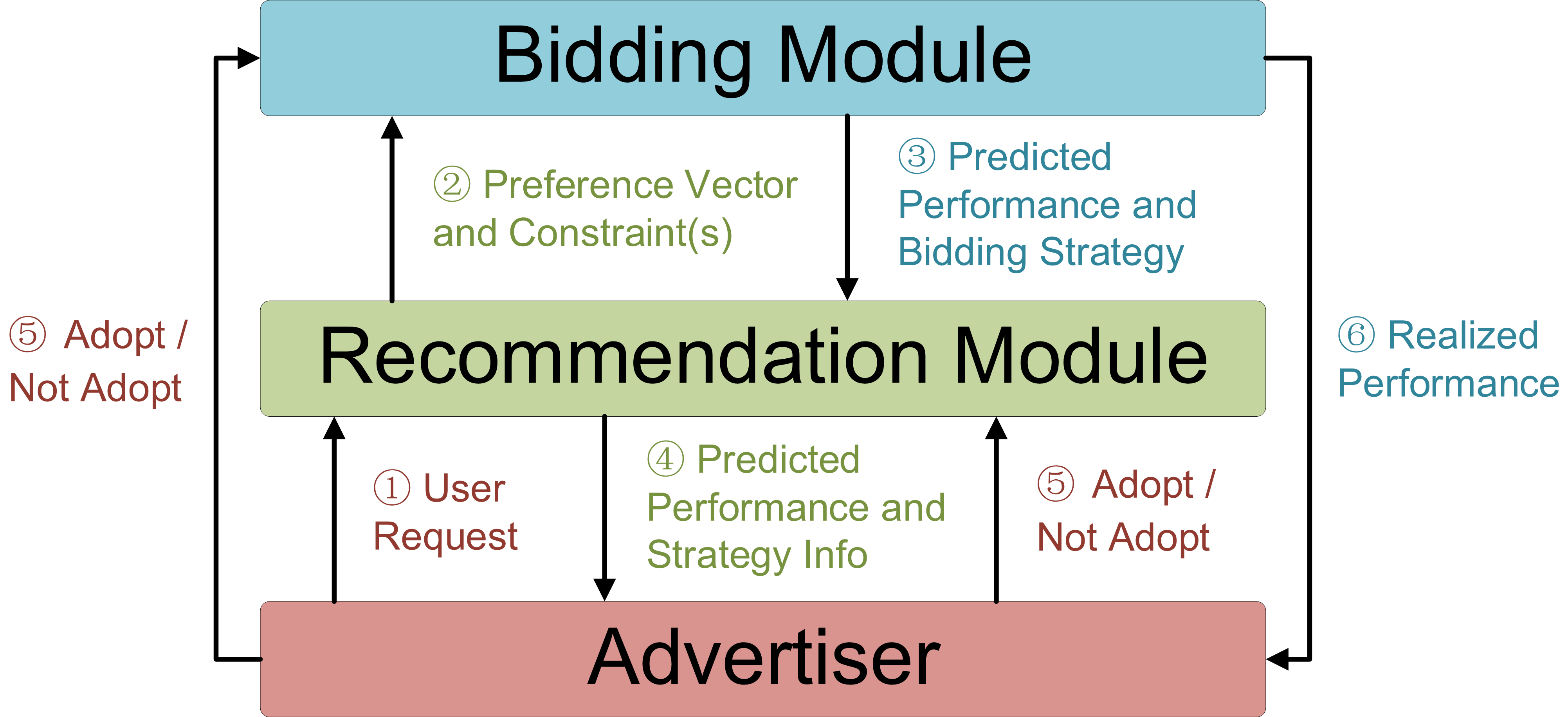}
  \caption
    {The architecture of an advertising strategy recommender system.}
\label{recommendation_system}
\end{figure}

To overcome the above obstacles, we design a new recommender system framework shown in Figure~\ref{recommendation_system} to learn about the advertisers' preference and then the optimization objective based on the interaction between advertisers and the recommender system. For each request of advertising strategy recommendation from advertisers, the recommender system processes as follows: 
(1) The \emph{recommendation module} receives the request from an advertiser and obtains her ad campaigns' information.
(2) The \emph{recommendation module} generates an estimated preference vector $\mathbf{w}$, which is sent, together with the constraint information such as the ad campaign's remaining budget, to the \emph{bidding module}. The \emph{recommendation module} inquires about the potential advertising performance for this preference vector. 
(3) The \emph{bidding module} generates the optimal bidding strategy $\Pi_{\mathbf{w}}$ based on the estimated preference vector and constraint(s), and returns the predicted advertising performance under this bidding strategy, $\mathbf{v}_{\Pi_{\mathbf{w}}} = [v_1, v_2, \cdots, v_n]^\top$ to the \emph{recommendation module}. (4) The \emph{recommendation module} displays the bidding strategy $\Pi_{\mathbf{w}}$ and the corresponding advertising performance $\mathbf{v}_{\Pi_{\mathbf{w}}}$ to the advertiser, which is easier to understand than only recommending the bidding strategy. (5) The advertiser chooses to adopt the recommending strategy or not by checking the advertising performance, and the \emph{recommendation module} collects the feedback from the advertiser. At the same time, once the advertiser adopts a recommendation, the \emph{bidding module} would conduct the real-time bidding algorithm based on the recommended strategy, to fulfill the promised advertising performance. (6) The \emph{bidding module} sends the realized advertising performance to the advertisers.


According to the above recommendation process, when the platform makes a bidding strategy recommendation to the advertiser, the advertiser would adopt the strategy if she is satisfied with the expected performance of the strategy; otherwise, the advertiser would reject. The adoption behaviors reflect advertisers' preferences and interests in the recommended advertising strategy and the associated performance. 
By efficiently learn the advertiser's preference, the platform can maximize the adoption rate. However, for a newly developed recommender system, a significant number of advertisers are likely to be entirely new without any historical adoption record, which is known as a \emph{cold-start} problem~\cite{park2006naive}. Thus, it is necessary to explore the advertisers' preferences in a continuous space for new advertisers. 
In the advertising platform, acquiring such information need to evaluate the advertisers' responses to different advertising strategies, which could be expensive and might reduce the advertiser's satisfaction in the short term. Hence, this raises the problem of balancing two competing goals: maximizing the advertiser's  satisfaction in the short term, and gathering information about advertiser's preference for the long-term development of the recommender system.

The above problem of the trade-off between two competing goals can be formulated as a feature-based exploitation/exploration process. Specifically, in this work, we 
leverage the technique from \emph{contextual bandit}, where an agent recommends the advertising strategy sequentially based on the contextual information of the ad campaigns while adapting its action selection strategy simultaneously based on the advertisers' feedback. We illustrate our modeling in detail in the next subsection.
\subsection{Contextual Bandit Modeling}

In this subsection, we formulate the contextual bandit problem to model the advertising strategy recommendation.
We formally define the notions of \emph{agent}, \emph{state}, \emph{action} and \emph{reward} in the context of strategy recommendation. The \emph{agent} is the strategy recommender system, and all advertisers visiting the system constitute the environment. When an  advertiser enters the recommender system, the agent needs to estimate an appropriate preference vector $\mathbf{w}$ for each of the advertiser’s ad campaigns based on the advertiser's information and historical behaviors of strategy adoption observed by the agent. With the advertiser's preference vector, we can derive the optimal bidding strategy and the corresponding advertising performance through the ad auction simulator. Then, for each ad campaign, the agent displays the recommended bidding strategy and estimated  advertising performance to the advertiser. The advertiser’s adoption behavior (adopt or reject the recommendation) can be regarded as the \emph{reward} of the environment to the agent. We summarize the state $\mathcal{S}$, action $\mathcal{A}$, and reward $\mathcal{R}$ of the contextual bandit problem as follows:

\begin{enumerate}

    \item
    Status $\mathcal{S}$: Information related to the ad campaign that the platform can observe, such as the ad campaign's information, the advertiser’s historical adoption-related information and etc. We use a feature vector $\mathbf{x}$ to represent the state $\mathcal{S}$.
    
    \item
    Action $\mathcal{A}$: Estimating the advertiser's preference vector $\mathbf{w}$. The agent obtains the (estimated) optimal bidding strategy and the corresponding advertising performance, by 
    sending the constraint information in the state $\mathbf{x}$ and the preference vector $\mathbf{w}$ to the bidding module.  
    
    \item
    Reward $\mathcal{R}$: We set the reward to be $1$ when the advertiser adopts the recommended advertising strategy; otherwise the reward is $0$. The expected reward $\mathcal{R}(\mathcal{S},\mathcal{A})$ represents the expected adoption rate under state $\mathcal{S}$ and action $\mathcal{A}$. That is the reward of the action is the advertiser's adoption feedback, and we need to maximize the reward by recommending the strategy most likely to be adopted by the advertiser.
\end{enumerate}

Based on the above modeling, the agent continuously recommends strategies for visiting advertisers. We number the discrete recommendation rounds as $t=1,2,3,...$. In the $t$-th round, the contextual bandit algorithm works as follows:

\begin{enumerate}

     \item
     The agent observes the feature vector $\mathbf{x}_t$ of the current ad campaign.
    
     \item
    Based on $\mathbf{x}_t$, the agent predicts a preference vector $\mathbf{w}_t$, and sends the constraint information in $\mathbf{x}_t$ together with $\mathbf{w}_t$ to the bidding module, which  performs the simulated bidding and returns the result $\mathbf{v}_t$. Then the agent recommends $\mathbf{v}_t$ to the advertiser and gets the reward $r_{t,\mathbf{w}_t}$.
    
    \item
     The observation of the current round $(\mathbf{x}_t,\mathbf{w}_t,\mathbf{v}_t,r_{t,\mathbf{w}_t})$ is used to help the agent make its future decision.
\end{enumerate}

In the above process, the \emph{T-trial Payoff} for $T$ rounds recommendations is $\sum_{t=1}^{T} r_{t,\mathbf{w}_t}$. Similarly, we define the \emph{Optimal Expected T-trial Payoff} as $\mathbf{E}\left[\sum_{t=1}^{T} r_{t, \mathbf{w}^{*}_t}\right]$, where the preference vector $\mathbf{w}^{*}_t$ is the true preference of the advertiser in the $t$-th round, and the optimal bidding strategy based on $\mathbf{w}^{*}_t$ would get the maximum advertising performance for advertisers. Our goal is to maximize the \emph{Expected T-trial Payoff} or equivalently, we can also regard the goal as minimizing the \emph{Expected T-trial Regret} $R(T)$ for $T$ rounds recommendations as follows:
\begin{equation}
\label{t_regret}
R(T) \quad \stackrel{\text{def}}{=} \quad \mathbf{E}\left[\sum_{t=1}^{T} r_{t, \mathbf{w}^{*}_t}\right]-\mathbf{E}\left[\sum_{t=1}^{T} r_{t, \mathbf{w}_t}\right].
\end{equation}

The \emph{Expected T-trial Regret} $R(T)$ can be interpreted as the gap between the expected adoption amount of the optimal recommendation strategy and our designed recommendation strategy under $T$ rounds of recommendations, and $\frac{R(T)}{T}$ is the gap between the expected adoption rate of the optimal recommendation strategy and the actual recommendation strategy.

\section{Algorithm Design}
\label{algorithm_design}
In the classic contextual bandit algorithm, the agent chooses to pull the arms based on some strategies to jointly learn the expected reward of each arm and the accumulated reward of the selected arms over a period. 
In the classic contextual bandit, the number of possible actions is usually discrete and finite, while the action space (the possible values of the preference vector) in advertising strategy recommendation is a high-dimensional continuous space. Furthermore, for each preference vector, it is also time consuming to solve a large scale of linear programming to obtain the bidding strategy and the  advertising performance in the ad auction simulator.  
With regard to this issue, we cannot output each action's estimated reward at the same time, which makes it difficult to directly apply the existing bandit algorithms such as $\epsilon$-greedy or upper confidence bound~\cite{kuleshov2014algorithms}. To solve the above problem, we divide  the reward learning process into two steps: (1) Choosing the action based on the observed information. (2) Establishing the relation between (the estimated bidding strategy and  the advertising performance) and the advertiser's adoption rate.

We first build the connection between the advertiser information and the advertising preference. More specifically, based on the action selection strategy, we obtain the preference vector $\mathbf{w}$ under the state $\mathcal{S}$ that the platform can observe as:
\begin{flalign}
\begin{aligned}
\label{neural_network}
    \mathbf{w}&=f(\textbf{x})\\
    &=f(advertiser\ behaviors, ad\ profile, scenario, etc.),
\end{aligned}
\end{flalign}
where the function $f$ is a mapping between the environmental state $S$ and the preference vector $\mathbf{w}$, the input $\textbf{x}$ of $f$ is the representation of the state $\mathcal{S}$, and the output of the network is $\mathbf{w}$. We use a multi-layer perceptron model (MLP) to build such a mapping. 
The label of the neural network (i.e. the reward of the action) is the adoption behavior of the advertiser. 
We recall that $\mathbf{v}$ is the advertising performance under the preference vector $\mathbf{w}$, and the value of $\mathbf{w}^\top \cdot \mathbf{v}$ is the utility that the advertiser can obtain. Since the advertiser’s adoption rate is positively correlated with $\mathbf{w}^\top \cdot \mathbf{v}$, we model the relation of $\mathbf{w}^\top \cdot \mathbf{v}$ and the adoption rate as follows:
\begin{equation}
\label{accept_rate}
p(\mathrm{Adopt})=\sigma(\mathbf{w}^\top \cdot \mathbf{v}),
\end{equation}
where $\sigma$ is the sigmoid function, and the  advertising performance $\mathbf{v}$ based on $\mathbf{w}$ is also part of the model input. In practice, advertising performance $\mathbf{v}$ could be normalized by the advertiser's budget since we are only interested in the relative value in $\mathbf{v}$ and $\mathbf{w}$.

Based on the above discussion, the estimation of the action value (the preference vector $\mathbf{w}$) of the network can be updated through gradient descent. For each iteration, the model first observes the environmental feature $\textbf{x}$ and estimates the preference $\mathbf{w}$, predicts the advertising performance $\mathbf{v}$ according to $\mathbf{w}$, and calculates the adoption rate $p(\textbf{x}, \mathbf{v})$. Then, we update the parameters of the model through the following loss function:
\begin{equation}
L=-\frac{1}{N}\sum_{(\textbf{x}, \mathbf{v}, y) \in \mathcal{D}}^{} (y \times log p(\textbf{x}, \mathbf{v}) + (1 - y) \times log (1 - p(\textbf{x}, \mathbf{v})),
\end{equation}
where the set $\mathcal{D}$ is the data set of size $N$ in this update iteration, and the label $y$ is the advertiser's realized adoption behavior.

\noindent\textbf{Action selection strategy.} In the context of advertising strategy recommendation, exploration means recommending strategies based on new possible preference vectors to explore the potential interests of advertisers, and exploitation means recommending corresponding strategies based on the  currently learned preference vector $\mathbf{w}$, which is the output of the model. Thompson Sampling is one of the efficient ways to make the trade-off between the exploitation and exploration~\cite{thompson1933likelihood}. Generally speaking, Thompson Sampling requires Bayesian processing of model parameters. In each step, Thompson Sampling samples a new set of model parameters, and then selects the preference vector based on the model with the sampling parameters. This can be seen as a kind of random hypothesis test: the more likely parameters are sampled more frequently and thus be rejected or adopted more quickly. Specifically, the process of Thompson Sampling is as follows:

\begin{enumerate}
     \item
     \label{first_step}
     Sampling a new set of parameters for the model.
     \item
     Selecting the preference vector with the highest expected reward based on the model with the sampling parameters.
     \item
     Updating the model and going back to~(\ref{first_step}).
\end{enumerate}

Performing Thompson Sampling in a neural network needs to describe the uncertainty of the network. Bayesian models~\cite{mackay1992practical, neal1995bayesian, blundell2015weight} offer a mathematically grounded framework to represent the uncertainty of the models, but usually, come with a prohibitive computational cost~\cite{gal2016dropout}. Dropout, a simple yet effective way to avoid neural networks from overfitting~\cite{srivastava2014dropout}, is proved to be a Bayesian approximation representing the uncertainty of the deep learning model~\cite{gal2016dropout}. Inspired by the above discussion, we apply Dropout in neural networks to balance exploration and exploitation in our recommendation problem.

\section{Experiments}

In this section, we first evaluate our prototype recommender system through online evaluations, and then build a simulation environment to conduct extensive simulated evaluations of the proposed advertising strategy recommendation methods. We summarize our experimental results as follows: (1) The online evaluations demonstrate the potential benefits of introducing advertising strategy recommendation toward advertisers. The recommender system not only optimizes the advertiser's advertising performance but also increases the revenue of the platform. (2) The designed neural network is effective in the advertiser adoption rate optimization, which depends on the accurate prediction of advertiser preferences. (3) The Dropout trick can efficiently weigh the pros and cons between exploiting the existing preference information and exploring the potential preferences of advertisers. (4) We verify the generalization ability of the bandit algorithm through ablation studies.

\subsection{Online Evaluations}
\label{systen_online_evaluation}

\begin{table}
\centering
\caption{The advertising performance of the prototype recommender system in the online A/B test.}
\label{tab:online_ab_test}
\begin{tabular}{cccc}
\toprule
Metric & Result & Metric & Result \\
\midrule
ARPU & $+1.2\%$ & RPM & $+0.8\%$ \\
Click Number& $+1.5\%$ & Payment Number & $+3.3\%$   \\
CTR & $+1.0\%$ &  Payment Amount & $+4.3\%$  \\
Cost & $+1.3\%$ &  CVR &  $+1.9\%$ \\
ROI & $+3.1\%$ & PV & $+0.5\%$ \\
\bottomrule
\end{tabular}
\end{table}

\noindent\textbf{System Implementation.} We have deployed our prototype recommender system mentioned in Section~\ref{sec:current_system} in the Taobao display advertising platform since February 2020. The detailed implementation of the system includes three main components: First, we maintain two individual databases for the bid optimization module and the targeted user optimization module, which store recommended items (e.g. targeted users and corresponding bids) for different ads. The recommended items in the two databases are updated daily. Second, we implement a ad auction simulator, which uses advertisers' bidding logs sampled from online auctions to predict the advertising performance using a specific bidding strategy with certain optimization objectives and constraints. Finally, we maintain a database that stores the realized advertising performance of the advertisers.

When an advertiser issues a request of strategy recommendation, the recommender system selects the recommended strategies based on the bandit algorithm, and predicts the advertising performance of these strategies using the ad auction simulator. Then the system ranks the recommended strategies based on some rules on the advertising performance and shows the top-K recommendations to the advertiser, together with the corresponding advertising performance. 
Once the advertiser adopts some recommendations, the system would carry out the adopted recommended strategy immediately.

\noindent\textbf{Evaluation Results.} In our online experiment between 2020-05-14 and 2020-05-27, we carefully select advertisers with the same consumption level based on their historical Average Revenue per User (ARPU), the daily cost of advertisers in our experiments constitutes about $30\%$ of the total daily cost of all the advertisers in the platform. It is worth to note that none of the advertisers in our experiment had used the recommender system before 2020-05-14. In the A/B test, half of the advertisers using the recommender system in the evaluation period form the test group, and the other half forms the control group. The average open rate of the bid optimization module and targeted user optimization module in the test group is $15.8\%$ and $25.4\%$ respectively, indicating that some advertisers are willing to use our recommender system. The average adoption rate of the advertisers is $4.5\%$ and $3.8\%$ respectively, for bid recommendation and targeted user recommendation. We find that advertisers would select and adopt a few  recommendations carefully and cautiously, revealing the opportunity to introduce a learning module for personalized recommendations. The ARPU in the test group increases by $1.2\%$ compared with the control group, showing that advertisers are satisfied with the recommended strategies and are willing to invest more in advertising, and thus increase the revenue of the platform. Moreover, as shown in Table~\ref{tab:online_ab_test}, the overall advertising performance of ad campaigns in the test group is better than those in the control group, which verifies that the recommender system can indeed improve advertisers' advertising performance.

It could be shown that the above experimental results verify the effectiveness of the prototype recommender system. However, some weaknesses still exist in the infant system: (1) Advertisers need to select the recommendations they want from a large number of recommendations. (2) The improvement of advertising performance for advertisers is not significant. Consequently, the prototype recommender system needs to be augmented through extensions.

\subsection{Simulation Settings}

Compared with machine learning in a standard supervised environment, evaluations of bandit algorithms are more difficult~\cite{li2010contextual}. Besides, bandit algorithms usually require a lot of interaction with the environment, making it costly to conduct experiments in a real advertising environment due to the large expenses of trial and error~\cite{shi2019virtual}. Therefore, most of the previous researches verify the effectiveness of the algorithm by building a simulation environment~\cite{ie2019slateq, hao2020dynamic}. In this section, we introduce the design of our simulation environment, which are the \emph{advertiser module} and \emph{bidding module} of the system architecture in Figure~\ref{recommendation_system}. The advertiser module simulates the interaction between the advertisers and the recommendation module in the real advertising system. The bidding module obtains the expected advertising performance based on ad campaigns' bidding logs.

\noindent\textbf{Bidding module:} The bidding module optimizes the objective function $\textbf{w}^\top \cdot \textbf{v}$ under the budget constraint. The advertising indicators in the simulation environment are PV, Click Number, and GMV. We use offline linear programming~\cite{zhang2016optimal} to derive the optimal bidding strategy and the corresponding advertising performance based on the ad campaign's bidding logs .

\noindent\textbf{Advertiser module:} 
In the simulation environment, each ad campaign has a budget $B$ and a preference vector $\mathbf{w}^{*}$. We generate the advertiser's preference vector $\mathbf{w}^{*}$ as $\mathbf{w}^{*}=g(\mathrm{Parameter\ Set})$, where $g$ is a function based on a parameter set, which represents the advertiser characteristics observed by the platform. Specifically, we design the following scheme: given the feature matrix composed of the $m$ typical (basic) preference vectors $A_{n \times m}=[\mathbf{w}_1;\mathbf{w}_2;.. .;\mathbf{w}_m]$ and the advertiser’s feature vector $\mathbf{c}=[c_1,c_2,...,c_m]^\top$, where $c_i$ means the weight of the $i$-th typical preference $\mathbf{w}_i$ in the feature matrix. The preference vector of the advertiser is obtained by $\mathbf{w}^{*}=A \cdot c$. In the simulation, the agent can observe the feature matrix $A$ and the feature vector $\mathbf{c}$. It is worth to note that other relations can also be designed and tested, but we cannot know the relation in the production environment in advance.

We next model the advertiser’s adoption behavior. We apply the conditional logit model that is widely used in user selection modeling~\cite{louviere2000stated} to simulate the adoption behaviors of advertisers. In the conditional logit model, the probability that user $i$ selects item $j$ in the recommended set $\Omega$ is $P(j \mid \Omega)=\frac{e ^{u\left(x_{ij}\right)}}{\sum_{\ell \in \Omega} e^{u\left(x_{i \ell}\right)}}$, where $u \left(x_{ij}\right)$ is the utility of the commodity $j$ to the user $i$. The utility of the user selecting a ``null'' commodity (not selecting any commodity) is usually expressed by a constant~\cite{ie2019slateq}. Therefore, we design the advertiser's probability to adopt the recommended strategy in the simulation environment as:
\begin{equation}
\label{simulation_accept}
    \Pr(\texttt{Adopt}) = \frac{e^{u(\mathbf{w}^{*}, \mathbf{v}, \mathbf{v}')}}{e^{u(\mathbf{w}^{*}, \mathbf{v}, \mathbf{v}')} + C},
\end{equation}
where the vector $\mathbf{w}^{*}$ represents the advertiser’s preference, and the vector $\mathbf{v}$ represents the advertising performance obtained through bidding according to the advertiser’s preference vector $\mathbf{w}^{*}$. The vector $\textbf{v}'$ represents the advertising performance obtained through bidding under the model's estimated preference vector $\mathbf{w}$, and the value $C$ represents the utility of advertiser’s rejection. The  $u(\mathbf{w}^{*}, \mathbf{v}, \mathbf{v}')$ represents the utility function of the advertiser’s adoption, we design $u(\mathbf{w}^{*}, \mathbf{v}, \mathbf{v}') = \alpha \times \frac{{\mathbf{w}^{*}}^\top \cdot \mathbf{v}}{{\mathbf {w}^{*}}^\top \cdot \mathbf{v}-{\mathbf{w}^{*}}^\top \cdot \mathbf{v}'}$, where $\alpha$ is the parameter for adjusting the overall adoption rate of advertisers in the simulation environment. In the utility function, the term $\frac{{\mathbf{w}^{*}}^\top \cdot \mathbf{v}}{{ \mathbf{w}^{*}}^\top \cdot \mathbf{v}-{\mathbf{w}^{*}}^\top \cdot \mathbf{v}'}$ indicates the reciprocal of $\frac{{\mathbf{w}^{*}}^\top \cdot \mathbf {v}-{\mathbf{w}^{*}}^\top \cdot \mathbf{v}'}{{\mathbf{w}^{*}}^\top \cdot \mathbf{v}}$, which is the relative utility gap between the advertising performance under  the advertiser's estimated preference ${\mathbf{w}}$ and true preference ${\mathbf{w}^{*}}$. To prevent division by zero, we set the denominator greater than or equal to a small positive number $\epsilon$.
It can be seen from ~(\ref{simulation_accept}) that when the utility of the recommended strategy is closer to the optimal utility, the advertiser's adoption rate is higher, which is convenient for us to compare the effectiveness of different models.

In the initialization phase of the advertiser module, several ad campaigns with corresponding preference vectors and budgets are generated following the above procedure, and the visit event of the recommender system is randomly triggered. The ad campaigns in the simulation environment share the historical bidding logs.

\noindent\textbf{Evaluation metrics and training parameters:} The optimization goal of the contextual bandit algorithm is the \emph{Expected T-trial Regret} in $T$ rounds of interactions shown in~(\ref{t_regret}): the accumulated expected regret is  $\mathbf{E}\left[\sum_{t=1}^{T} r_{t, \mathbf{w}^{*}_t}\right]-\mathbf{E}\left[\sum_{t=1}^{T} r_{t, \mathbf{w}_t}\right]$. The accumulated adoption rate is $\frac{\sum_{t=1}^{T} r_{t,\mathbf{w}_t}}{T}$. We can use these two metrics to evaluate the performance of the model.

In the simulation experiments, input features of the model are preference-related features and advertiser's adoption behaviors related to the ad campaign. For preference-related features, we concatenate them together as one of the inputs. For advertiser's adoption behaviors, we apply average pooling to the corresponding model output $\mathbf{w}$ in historical adoption behaviors related to the ad. We use mini-batch SGD to train the model when it interacts with the environment, and use Adam~\cite{kingma2014adam} as the optimizer. To prevent the imbalance of the positive and negative samples to affect the model performance, we set the ratio of positive and negative samples in each training batch to $1:1$.

\subsection{Experimental Results}
\noindent\textbf{Exploration of advertiser's advertising performance space:} First, we briefly explore the advertising performance space of the advertisers in the Taobao online advertising environment. In this experiment, we select three types of typical objectives of advertisers in the platform: maximizing total impressions, maximizing total clicks, and maximizing GMV. Suppose the advertising performance indicators are PV, Click Number and GMV, the preference vectors of these three types of advertisers are $[1, 0, 0]^\top$, $[0, 1, 0]^\top$ and $[0, 0, 1]^\top$, respectively. We conduct the optimal bidding strategy under the budget constraint for the above three preference vectors in the ad auction simulator. The experimental results are shown in Table~\ref{tab:different_preference_result}, it can be seen that the performance of indicators to be optimized has been significantly improved under the corresponding preference vectors, which demonstrates the importance of understanding the advertisers’ preferences in advertising performance optimization.
\begin{table}
\centering
\caption{The advertising performance of the same ad campaign under different preferences. The values of each column are normalized by the maximum value in each column.}
\label{tab:different_preference_result}
\begin{tabular}{lccc}
\toprule
Objective Type & PV & Click Number & GMV \\
\midrule
Optimizing PV & \textbf{1.0000}  & 0.7747 &  0.6751 \\
Optimizing Click Number & 0.7825      & \textbf{1.0000} &  0.8269 \\
Optimizing GMV & 0.7377      & 0.8454     & \textbf{1.0000} \\
\bottomrule
\end{tabular}
\end{table}

\begin{table}
\centering
\caption{Comparisons of different models. The values of each column are normalized by the maximum value in each column. Metric 1 indicates accumulated expected regret, and Metric 2 indicates accumulated adoption rate.}
\label{tab:experiment_result}
\begin{tabular}{lcc}
\toprule
Model & Metric 1  &  Metric 2 \\
\midrule
Random Preference (No Learning Module)  & 1.0000  &  0.7606  \\
No Dropout  & 0.7429  &  0.8239  \\
Dropout ($40\%$, No Preference-Related Info)  & 0.7195  &  0.8535  \\
Dropout ($80\%$)  &  0.5384 &  0.9144  \\
Dropout ($20\%$)  &  0.4154 &  0.9725  \\
Dropout ($60\%$)  &  0.4078 &  0.9817  \\
Dropout ($40\%$)  & \textbf{0.3881}  & \textbf{1.0000}    \\
\bottomrule
\end{tabular}
\end{table}

\noindent\textbf{Comparison experimental results:} We show the effectiveness of the proposed contextual bandit algorithm through comparison experiments, in which we compare models with different Dropout ratios or without Dropout. We also implement a recommendation strategy with a random preference. In each experiment, the agent interacts with the environment for $2000$ rounds, and updates accumulated expected regret and accumulated adoption rate periodically. We show the experimental results in Table~\ref{tab:experiment_result}. 
From Table~\ref{tab:experiment_result}, we observe that the recommendation with a random preference causes huge degradation in the adoption rate, which motivates the necessity of considering advertiser preferences when recommending strategies. As shown in Table~\ref{tab:experiment_result}, even without using the Dropout trick, a model that explicitly learns the advertiser's preference can reduce the accumulated expected regret by $25.71\%$ compared to the recommendation strategy without a learning module. From Table~\ref{tab:experiment_result}, we also learn that the context bandit algorithms which apply Dropout are more effective than the algorithm without Dropout. We observe that as the Dropout ratio increases (from $20\%$, $40\%$, $60\%$ to $80\%$), the performance of the model first increases and then decreases. This is because (1) when the Dropout ratio is low, the model adopts a conservative exploration strategy. (2) when the dropout ratio is high, the model frequently explores the action space, which cannot make full use of the learned knowledge, resulting in performance degradation.

\begin{figure}[t]
\centering
    \begin{subfigure}{0.235\textwidth}
    \centering
        \includegraphics[width=1.0\linewidth]{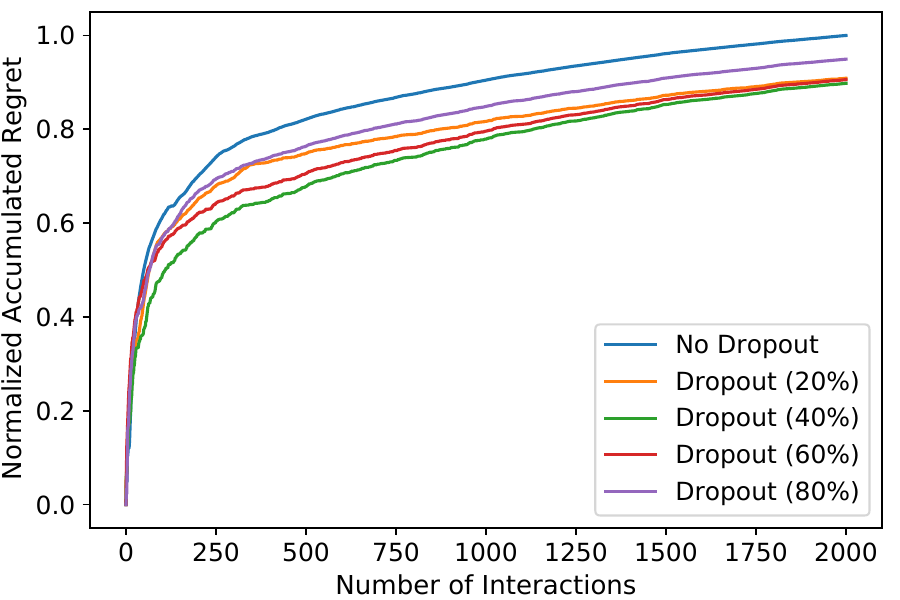}
        \caption{Accumulated expected regret.}
        \label{fig:regret_comparison}
    \end{subfigure}%
    \begin{subfigure}{0.235\textwidth}
    \centering
        \includegraphics[width=1.0\linewidth]{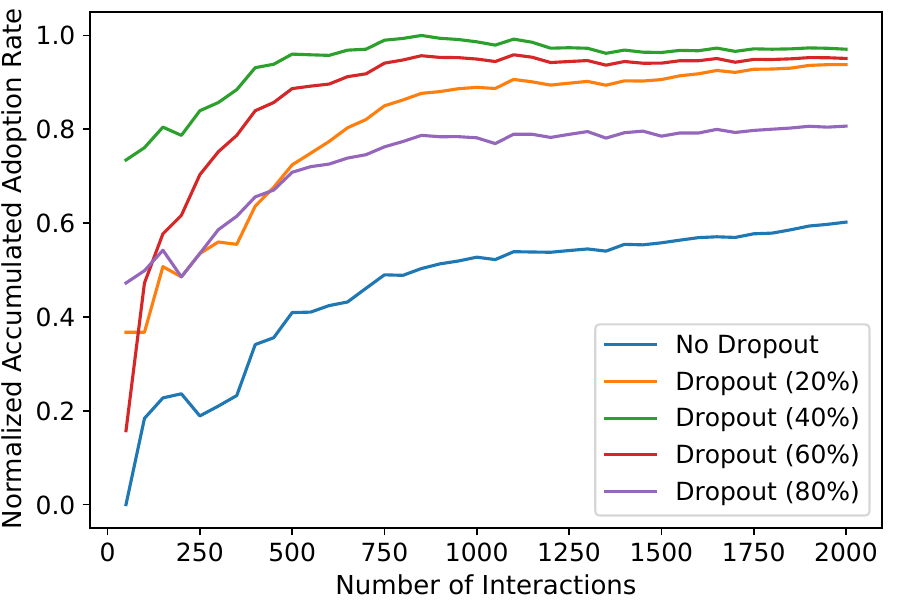}
        \caption{Accumulated adoption rate.}
        \label{fig:accept_comparison}
    \end{subfigure}
\caption{Comparisons of models with different Dropout ratios.}
\label{fig:comparison_experiment}
\end{figure}

In Figure~\ref{fig:comparison_experiment}, we show the curves of accumulated expected regret and accumulated adoption rate for the models with different Dropout ratios in various numbers of interactions. To better evaluate the performance difference after the model converges, we normalize the accumulated regret by using the function $y=log(x+1)$. From Figure~\ref{fig:comparison_experiment}, we observe that different models converge to different local optimums, and the growth speed of accumulated expected regret in all models decreases at first and then converges, and the accumulated adoption rate increases gradually and then converges. All these models learn the preferences of advertisers to some extent to improve the performance of the recommender system, compared with the recommendation strategy without a learning module.

\begin{figure}[t]
\centering
    \begin{subfigure}{0.235\textwidth}
    \centering
        \includegraphics[width=1.0\linewidth]{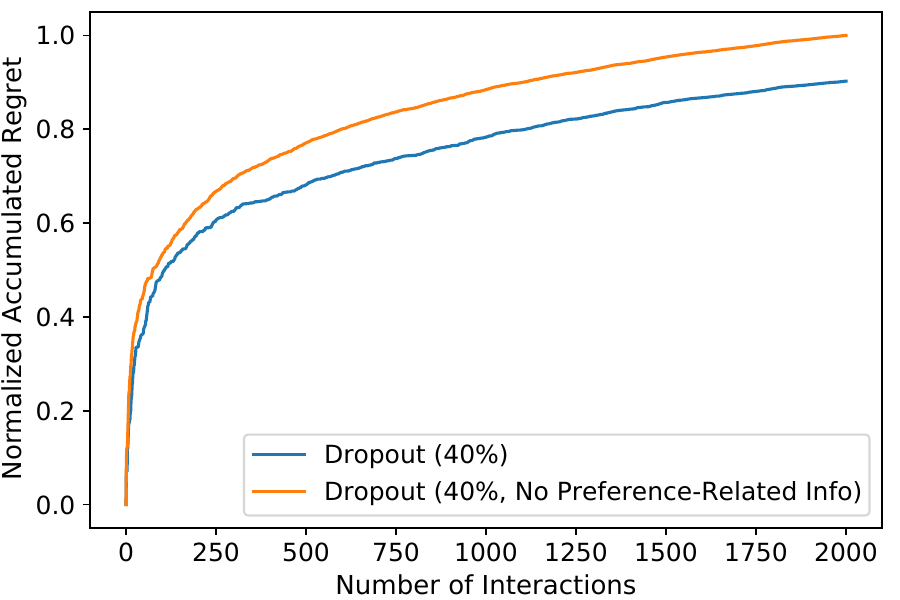}
        \caption{Accumulated expected regret.}
        \label{fig:regret_comparison_context}
    \end{subfigure}%
    \begin{subfigure}{0.235\textwidth}
    \centering
        \includegraphics[width=1.0\linewidth]{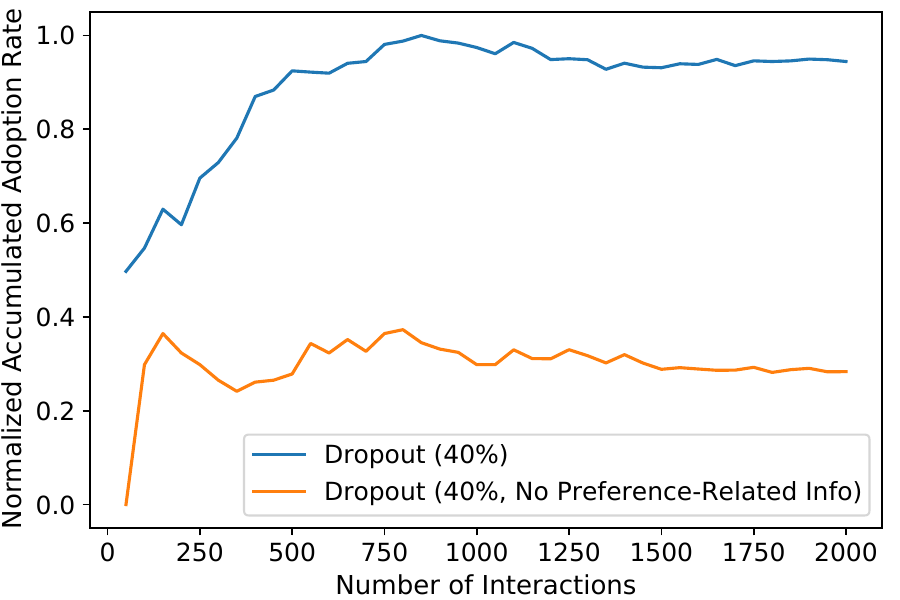}
        \caption{Accumulated adoption rate.}
        \label{fig:accept_comparison_context}
    \end{subfigure}
\caption{Impact of preference-related information.}
\label{fig:context_experiment}
\end{figure}

To verify the generalization ability of the model, we conduct a controlled experiment, in which the experimental group is a model with a Dropout ratio of $40\%$, and the control group is the same model without the preference-related information (only historical adoption information can be observed by the model). We draw the results in Figure~\ref{fig:context_experiment}, and the experimental results show that the performance of the model with the preference-related information is better than the model without it in both accumulated expected regret and accumulated adoption rate. This indicates that the model can learn the preferences of advertisers through understanding information related to $\mathbf{w}$, and thus verifying the generalization performance of the model.

\section{Conclusion}
In this work, we have investigated the problem of advertising strategy recommendations for online advertising. Through an online A/B test on a prototype recommender system, we demonstrate the potential benefits of recommending strategies to advertisers. We further augment the prototype system by recommending advertising strategy associated with the predicted advertising performance. We could exploit the advertisers' adoption behaviors to recommended strategies to learn advertiser's preference and optimize strategy adoption rate. We formulate the problem of strategy adoption rate maximization as a contextual bandit problem. 
The exploration is to learn the advertisers' preferences and optimization objectives, and the exploitation is to improve the adoption rate using the learned knowledge. We used Dropout trick to balance the exploration and exploitation. Experiments in a simulation environment verified the effectiveness of the system in the task of adoption rate optimization.

\bibliographystyle{ACM-Reference-Format}
\balance
\bibliography{sample-base}


\begin{thebibliography}{43}


\ifx \showCODEN    \undefined \def \showCODEN     #1{\unskip}     \fi
\ifx \showDOI      \undefined \def \showDOI       #1{#1}\fi
\ifx \showISBNx    \undefined \def \showISBNx     #1{\unskip}     \fi
\ifx \showISBNxiii \undefined \def \showISBNxiii  #1{\unskip}     \fi
\ifx \showISSN     \undefined \def \showISSN      #1{\unskip}     \fi
\ifx \showLCCN     \undefined \def \showLCCN      #1{\unskip}     \fi
\ifx \shownote     \undefined \def \shownote      #1{#1}          \fi
\ifx \showarticletitle \undefined \def \showarticletitle #1{#1}   \fi
\ifx \showURL      \undefined \def \showURL       {\relax}        \fi
\providecommand\bibfield[2]{#2}
\providecommand\bibinfo[2]{#2}
\providecommand\natexlab[1]{#1}
\providecommand\showeprint[2][]{arXiv:#2}

\bibitem[\protect\citeauthoryear{Basilico and Hofmann}{Basilico and
  Hofmann}{2004}]%
        {basilico2004unifying}
\bibfield{author}{\bibinfo{person}{Justin Basilico} {and}
  \bibinfo{person}{Thomas Hofmann}.} \bibinfo{year}{2004}\natexlab{}.
\newblock \showarticletitle{Unifying collaborative and content-based
  filtering}. In \bibinfo{booktitle}{\emph{ICML}}. \bibinfo{pages}{9}.
\newblock


\bibitem[\protect\citeauthoryear{Blundell, Cornebise, Kavukcuoglu, and
  Wierstra}{Blundell et~al\mbox{.}}{2015}]%
        {blundell2015weight}
\bibfield{author}{\bibinfo{person}{Charles Blundell}, \bibinfo{person}{Julien
  Cornebise}, \bibinfo{person}{Koray Kavukcuoglu}, {and} \bibinfo{person}{Daan
  Wierstra}.} \bibinfo{year}{2015}\natexlab{}.
\newblock \showarticletitle{Weight uncertainty in neural networks}. In
  \bibinfo{booktitle}{\emph{ICML}}. \bibinfo{pages}{1613--1622}.
\newblock


\bibitem[\protect\citeauthoryear{Cai, Ren, Zhang, Malialis, Wang, Yu, and
  Guo}{Cai et~al\mbox{.}}{2017}]%
        {cai2017real}
\bibfield{author}{\bibinfo{person}{Han Cai}, \bibinfo{person}{Kan Ren},
  \bibinfo{person}{Weinan Zhang}, \bibinfo{person}{Kleanthis Malialis},
  \bibinfo{person}{Jun Wang}, \bibinfo{person}{Yong Yu}, {and}
  \bibinfo{person}{Defeng Guo}.} \bibinfo{year}{2017}\natexlab{}.
\newblock \showarticletitle{Real-time bidding by reinforcement learning in
  display advertising}. In \bibinfo{booktitle}{\emph{WSDM}}.
  \bibinfo{pages}{661--670}.
\newblock


\bibitem[\protect\citeauthoryear{Cheng, Koc, Harmsen, Shaked, Chandra, Aradhye,
  Anderson, Corrado, Chai, Ispir, et~al\mbox{.}}{Cheng et~al\mbox{.}}{2016}]%
        {cheng2016wide}
\bibfield{author}{\bibinfo{person}{Heng-Tze Cheng}, \bibinfo{person}{Levent
  Koc}, \bibinfo{person}{Jeremiah Harmsen}, \bibinfo{person}{Tal Shaked},
  \bibinfo{person}{Tushar Chandra}, \bibinfo{person}{Hrishi Aradhye},
  \bibinfo{person}{Glen Anderson}, \bibinfo{person}{Greg Corrado},
  \bibinfo{person}{Wei Chai}, \bibinfo{person}{Mustafa Ispir}, {et~al\mbox{.}}}
  \bibinfo{year}{2016}\natexlab{}.
\newblock \showarticletitle{Wide \& deep learning for recommender systems}. In
  \bibinfo{booktitle}{\emph{DLRS}}. \bibinfo{pages}{7--10}.
\newblock


\bibitem[\protect\citeauthoryear{Covington, Adams, and Sargin}{Covington
  et~al\mbox{.}}{2016}]%
        {covington2016youtube}
\bibfield{author}{\bibinfo{person}{Paul Covington}, \bibinfo{person}{Jay
  Adams}, {and} \bibinfo{person}{Emre Sargin}.}
  \bibinfo{year}{2016}\natexlab{}.
\newblock \showarticletitle{Deep neural networks for youtube recommendations}.
  In \bibinfo{booktitle}{\emph{RecSys}}. \bibinfo{pages}{191--198}.
\newblock


\bibitem[\protect\citeauthoryear{Evans}{Evans}{2008}]%
        {evans2008economics}
\bibfield{author}{\bibinfo{person}{David~S Evans}.}
  \bibinfo{year}{2008}\natexlab{}.
\newblock \showarticletitle{The economics of the online advertising industry}.
\newblock \bibinfo{journal}{\emph{Review of Network Economics}}
  \bibinfo{volume}{7}, \bibinfo{number}{3} (\bibinfo{year}{2008}),
  \bibinfo{pages}{1--33}.
\newblock


\bibitem[\protect\citeauthoryear{Evans}{Evans}{2009}]%
        {evans2009online}
\bibfield{author}{\bibinfo{person}{David~S Evans}.}
  \bibinfo{year}{2009}\natexlab{}.
\newblock \showarticletitle{The online advertising industry: Economics,
  evolution, and privacy}.
\newblock \bibinfo{journal}{\emph{Journal of Economic Perspectives}}
  \bibinfo{volume}{23}, \bibinfo{number}{3} (\bibinfo{year}{2009}),
  \bibinfo{pages}{37--60}.
\newblock


\bibitem[\protect\citeauthoryear{Gal and Ghahramani}{Gal and
  Ghahramani}{2016}]%
        {gal2016dropout}
\bibfield{author}{\bibinfo{person}{Y Gal} {and} \bibinfo{person}{Z
  Ghahramani}.} \bibinfo{year}{2016}\natexlab{}.
\newblock \showarticletitle{Dropout as a Bayesian approximation: Representing
  model uncertainty in deep learning}. In \bibinfo{booktitle}{\emph{ICML}}.
  \bibinfo{pages}{1651--1660}.
\newblock


\bibitem[\protect\citeauthoryear{Goldfarb and Tucker}{Goldfarb and
  Tucker}{2011}]%
        {goldfarb2011online}
\bibfield{author}{\bibinfo{person}{Avi Goldfarb} {and}
  \bibinfo{person}{Catherine Tucker}.} \bibinfo{year}{2011}\natexlab{}.
\newblock \showarticletitle{Online display advertising: Targeting and
  obtrusiveness}.
\newblock \bibinfo{journal}{\emph{Marketing Science}} \bibinfo{volume}{30},
  \bibinfo{number}{3} (\bibinfo{year}{2011}), \bibinfo{pages}{389--404}.
\newblock


\bibitem[\protect\citeauthoryear{Guo, Tang, Ye, Li, and He}{Guo
  et~al\mbox{.}}{2017}]%
        {guo2017deepfm}
\bibfield{author}{\bibinfo{person}{Huifeng Guo}, \bibinfo{person}{Ruiming
  Tang}, \bibinfo{person}{Yunming Ye}, \bibinfo{person}{Zhenguo Li}, {and}
  \bibinfo{person}{Xiuqiang He}.} \bibinfo{year}{2017}\natexlab{}.
\newblock \showarticletitle{DeepFM: a factorization-machine based neural
  network for CTR prediction}. In \bibinfo{booktitle}{\emph{IJCAI}}.
  \bibinfo{pages}{1725--1731}.
\newblock


\bibitem[\protect\citeauthoryear{Guo, Lu, Zhang, Jin, Zheng, Wu, Li, Xu, Li,
  Lu, et~al\mbox{.}}{Guo et~al\mbox{.}}{2020}]%
        {guo2020deep}
\bibfield{author}{\bibinfo{person}{Liyi Guo}, \bibinfo{person}{Rui Lu},
  \bibinfo{person}{Haoqi Zhang}, \bibinfo{person}{Junqi Jin},
  \bibinfo{person}{Zhenzhe Zheng}, \bibinfo{person}{Fan Wu},
  \bibinfo{person}{Jin Li}, \bibinfo{person}{Haiyang Xu}, \bibinfo{person}{Han
  Li}, \bibinfo{person}{Wenkai Lu}, {et~al\mbox{.}}}
  \bibinfo{year}{2020}\natexlab{}.
\newblock \showarticletitle{A Deep Prediction Network for Understanding
  Advertiser Intent and Satisfaction}. In \bibinfo{booktitle}{\emph{CIKM}}.
  \bibinfo{pages}{2501--2508}.
\newblock


\bibitem[\protect\citeauthoryear{Hao, Peng, Ma, Wang, Jin, Hao, Chen, Bai, Xie,
  Xu, et~al\mbox{.}}{Hao et~al\mbox{.}}{2020}]%
        {hao2020dynamic}
\bibfield{author}{\bibinfo{person}{Xiaotian Hao}, \bibinfo{person}{Zhaoqing
  Peng}, \bibinfo{person}{Yi Ma}, \bibinfo{person}{Guan Wang},
  \bibinfo{person}{Junqi Jin}, \bibinfo{person}{Jianye Hao},
  \bibinfo{person}{Shan Chen}, \bibinfo{person}{Rongquan Bai},
  \bibinfo{person}{Mingzhou Xie}, \bibinfo{person}{Miao Xu}, {et~al\mbox{.}}}
  \bibinfo{year}{2020}\natexlab{}.
\newblock \showarticletitle{Dynamic knapsack optimization towards efficient
  multi-channel sequential advertising}. In \bibinfo{booktitle}{\emph{ICML}}.
  \bibinfo{pages}{4060--4070}.
\newblock


\bibitem[\protect\citeauthoryear{Ie, Jain, Wang, Narvekar, Agarwal, Wu, Cheng,
  Chandra, and Boutilier}{Ie et~al\mbox{.}}{2019}]%
        {ie2019slateq}
\bibfield{author}{\bibinfo{person}{Eugene Ie}, \bibinfo{person}{Vihan Jain},
  \bibinfo{person}{Jing Wang}, \bibinfo{person}{Sanmit Narvekar},
  \bibinfo{person}{Ritesh Agarwal}, \bibinfo{person}{Rui Wu},
  \bibinfo{person}{Heng-Tze Cheng}, \bibinfo{person}{Tushar Chandra}, {and}
  \bibinfo{person}{Craig Boutilier}.} \bibinfo{year}{2019}\natexlab{}.
\newblock \showarticletitle{SLATEQ: a tractable decomposition for reinforcement
  learning with recommendation sets}. In \bibinfo{booktitle}{\emph{IJCAI}}.
  \bibinfo{pages}{2592--2599}.
\newblock


\bibitem[\protect\citeauthoryear{Kingma and Ba}{Kingma and Ba}{2014}]%
        {kingma2014adam}
\bibfield{author}{\bibinfo{person}{Diederik~P Kingma} {and}
  \bibinfo{person}{Jimmy Ba}.} \bibinfo{year}{2014}\natexlab{}.
\newblock \showarticletitle{Adam: A method for stochastic optimization}.
\newblock \bibinfo{journal}{\emph{arXiv preprint arXiv:1412.6980}}
  (\bibinfo{year}{2014}).
\newblock


\bibitem[\protect\citeauthoryear{Kitts, Krishnan, Yadav, Zeng, Badeau, Potter,
  Tolkachov, Thornburg, and Janga}{Kitts et~al\mbox{.}}{2017}]%
        {kitts2017ad}
\bibfield{author}{\bibinfo{person}{Brendan Kitts}, \bibinfo{person}{Michael
  Krishnan}, \bibinfo{person}{Ishadutta Yadav}, \bibinfo{person}{Yongbo Zeng},
  \bibinfo{person}{Garrett Badeau}, \bibinfo{person}{Andrew Potter},
  \bibinfo{person}{Sergey Tolkachov}, \bibinfo{person}{Ethan Thornburg}, {and}
  \bibinfo{person}{Satyanarayana~Reddy Janga}.}
  \bibinfo{year}{2017}\natexlab{}.
\newblock \showarticletitle{Ad Serving with Multiple KPIs}. In
  \bibinfo{booktitle}{\emph{SIGKDD}}. \bibinfo{pages}{1853--1861}.
\newblock


\bibitem[\protect\citeauthoryear{Kuleshov and Precup}{Kuleshov and
  Precup}{2014}]%
        {kuleshov2014algorithms}
\bibfield{author}{\bibinfo{person}{Volodymyr Kuleshov} {and}
  \bibinfo{person}{Doina Precup}.} \bibinfo{year}{2014}\natexlab{}.
\newblock \showarticletitle{Algorithms for multi-armed bandit problems}.
\newblock \bibinfo{journal}{\emph{arXiv preprint arXiv:1402.6028}}
  (\bibinfo{year}{2014}).
\newblock


\bibitem[\protect\citeauthoryear{Lahaie, Pennock, Saberi, and Vohra}{Lahaie
  et~al\mbox{.}}{2007}]%
        {lahaie2007sponsored}
\bibfield{author}{\bibinfo{person}{S{\'e}bastien Lahaie},
  \bibinfo{person}{David~M Pennock}, \bibinfo{person}{Amin Saberi}, {and}
  \bibinfo{person}{Rakesh~V Vohra}.} \bibinfo{year}{2007}\natexlab{}.
\newblock \showarticletitle{Sponsored search auctions}.
\newblock \bibinfo{journal}{\emph{Algorithmic Game Theory}}
  \bibinfo{volume}{1} (\bibinfo{year}{2007}), \bibinfo{pages}{699--716}.
\newblock


\bibitem[\protect\citeauthoryear{Li, Chu, Langford, and Schapire}{Li
  et~al\mbox{.}}{2010}]%
        {li2010contextual}
\bibfield{author}{\bibinfo{person}{Lihong Li}, \bibinfo{person}{Wei Chu},
  \bibinfo{person}{John Langford}, {and} \bibinfo{person}{Robert~E Schapire}.}
  \bibinfo{year}{2010}\natexlab{}.
\newblock \showarticletitle{A contextual-bandit approach to personalized news
  article recommendation}. In \bibinfo{booktitle}{\emph{WWW}}.
  \bibinfo{pages}{661--670}.
\newblock


\bibitem[\protect\citeauthoryear{Linden, Smith, and York}{Linden
  et~al\mbox{.}}{2003}]%
        {linden2003amazon}
\bibfield{author}{\bibinfo{person}{Greg Linden}, \bibinfo{person}{Brent Smith},
  {and} \bibinfo{person}{Jeremy York}.} \bibinfo{year}{2003}\natexlab{}.
\newblock \showarticletitle{Amazon.com recommendations: Item-to-item
  collaborative filtering}.
\newblock \bibinfo{journal}{\emph{IEEE Internet Computing}}
  \bibinfo{volume}{7}, \bibinfo{number}{1} (\bibinfo{year}{2003}),
  \bibinfo{pages}{76--80}.
\newblock


\bibitem[\protect\citeauthoryear{Lops, De~Gemmis, and Semeraro}{Lops
  et~al\mbox{.}}{2011}]%
        {lops2011content}
\bibfield{author}{\bibinfo{person}{Pasquale Lops}, \bibinfo{person}{Marco
  De~Gemmis}, {and} \bibinfo{person}{Giovanni Semeraro}.}
  \bibinfo{year}{2011}\natexlab{}.
\newblock \showarticletitle{Content-based recommender systems: State of the art
  and trends}.
\newblock In \bibinfo{booktitle}{\emph{Recommender Systems Handbook}}.
  \bibinfo{pages}{73--105}.
\newblock


\bibitem[\protect\citeauthoryear{Louviere, Hensher, and Swait}{Louviere
  et~al\mbox{.}}{2000}]%
        {louviere2000stated}
\bibfield{author}{\bibinfo{person}{Jordan~J Louviere}, \bibinfo{person}{David~A
  Hensher}, {and} \bibinfo{person}{Joffre~D Swait}.}
  \bibinfo{year}{2000}\natexlab{}.
\newblock \bibinfo{booktitle}{\emph{Stated choice methods: analysis and
  applications}}.
\newblock \bibinfo{publisher}{Cambridge University Press}.
\newblock


\bibitem[\protect\citeauthoryear{Lu, Wu, Mao, Wang, and Zhang}{Lu
  et~al\mbox{.}}{2015}]%
        {lu2015recommender}
\bibfield{author}{\bibinfo{person}{Jie Lu}, \bibinfo{person}{Dianshuang Wu},
  \bibinfo{person}{Mingsong Mao}, \bibinfo{person}{Wei Wang}, {and}
  \bibinfo{person}{Guangquan Zhang}.} \bibinfo{year}{2015}\natexlab{}.
\newblock \showarticletitle{Recommender system application developments: a
  survey}.
\newblock \bibinfo{journal}{\emph{Decision Support Systems}}
  \bibinfo{volume}{74} (\bibinfo{year}{2015}), \bibinfo{pages}{12--32}.
\newblock


\bibitem[\protect\citeauthoryear{MacKay}{MacKay}{1992}]%
        {mackay1992practical}
\bibfield{author}{\bibinfo{person}{David~JC MacKay}.}
  \bibinfo{year}{1992}\natexlab{}.
\newblock \showarticletitle{A practical Bayesian framework for backpropagation
  networks}.
\newblock \bibinfo{journal}{\emph{Neural Computation}} \bibinfo{volume}{4},
  \bibinfo{number}{3} (\bibinfo{year}{1992}), \bibinfo{pages}{448--472}.
\newblock


\bibitem[\protect\citeauthoryear{Maehara, Narita, Baba, and Kawabata}{Maehara
  et~al\mbox{.}}{2018}]%
        {maehara2018optimal}
\bibfield{author}{\bibinfo{person}{Takanori Maehara}, \bibinfo{person}{Atsuhiro
  Narita}, \bibinfo{person}{Jun Baba}, {and} \bibinfo{person}{Takayuki
  Kawabata}.} \bibinfo{year}{2018}\natexlab{}.
\newblock \showarticletitle{Optimal bidding strategy for brand advertising}. In
  \bibinfo{booktitle}{\emph{IJCAI}}. \bibinfo{pages}{424--432}.
\newblock


\bibitem[\protect\citeauthoryear{Neal}{Neal}{1995}]%
        {neal1995bayesian}
\bibfield{author}{\bibinfo{person}{Radford~M Neal}.}
  \bibinfo{year}{1995}\natexlab{}.
\newblock \emph{\bibinfo{title}{Bayesian learning for neural networks}}.
\newblock \bibinfo{thesistype}{Ph.D. Dissertation}. \bibinfo{school}{University
  of Toronto}.
\newblock


\bibitem[\protect\citeauthoryear{Park, Pennock, Madani, Good, and DeCoste}{Park
  et~al\mbox{.}}{2006}]%
        {park2006naive}
\bibfield{author}{\bibinfo{person}{Seung-Taek Park}, \bibinfo{person}{David
  Pennock}, \bibinfo{person}{Omid Madani}, \bibinfo{person}{Nathan Good}, {and}
  \bibinfo{person}{Dennis DeCoste}.} \bibinfo{year}{2006}\natexlab{}.
\newblock \showarticletitle{Na{\"\i}ve filterbots for robust cold-start
  recommendations}. In \bibinfo{booktitle}{\emph{SIGKDD}}.
  \bibinfo{pages}{699--705}.
\newblock


\bibitem[\protect\citeauthoryear{Pazzani and Billsus}{Pazzani and
  Billsus}{2007}]%
        {pazzani2007content}
\bibfield{author}{\bibinfo{person}{Michael~J Pazzani} {and}
  \bibinfo{person}{Daniel Billsus}.} \bibinfo{year}{2007}\natexlab{}.
\newblock \showarticletitle{Content-based recommendation systems}.
\newblock In \bibinfo{booktitle}{\emph{The Adaptive Web}}.
  \bibinfo{pages}{325--341}.
\newblock


\bibitem[\protect\citeauthoryear{Ren, Zhang, Chang, Rong, Yu, and Wang}{Ren
  et~al\mbox{.}}{2017}]%
        {ren2017bidding}
\bibfield{author}{\bibinfo{person}{Kan Ren}, \bibinfo{person}{Weinan Zhang},
  \bibinfo{person}{Ke Chang}, \bibinfo{person}{Yifei Rong},
  \bibinfo{person}{Yong Yu}, {and} \bibinfo{person}{Jun Wang}.}
  \bibinfo{year}{2017}\natexlab{}.
\newblock \showarticletitle{Bidding machine: Learning to bid for directly
  optimizing profits in display advertising}.
\newblock \bibinfo{journal}{\emph{IEEE Transactions on Knowledge and Data
  Engineering}} \bibinfo{volume}{30}, \bibinfo{number}{4}
  (\bibinfo{year}{2017}), \bibinfo{pages}{645--659}.
\newblock


\bibitem[\protect\citeauthoryear{Resnick, Zeckhauser, Swanson, and
  Lockwood}{Resnick et~al\mbox{.}}{2006}]%
        {resnick2006value}
\bibfield{author}{\bibinfo{person}{Paul Resnick}, \bibinfo{person}{Richard
  Zeckhauser}, \bibinfo{person}{John Swanson}, {and} \bibinfo{person}{Kate
  Lockwood}.} \bibinfo{year}{2006}\natexlab{}.
\newblock \showarticletitle{The value of reputation on eBay: A controlled
  experiment}.
\newblock \bibinfo{journal}{\emph{Experimental Economics}} \bibinfo{volume}{9},
  \bibinfo{number}{2} (\bibinfo{year}{2006}), \bibinfo{pages}{79--101}.
\newblock


\bibitem[\protect\citeauthoryear{Sarwar, Karypis, Konstan, and Riedl}{Sarwar
  et~al\mbox{.}}{2001}]%
        {sarwar2001item}
\bibfield{author}{\bibinfo{person}{Badrul Sarwar}, \bibinfo{person}{George
  Karypis}, \bibinfo{person}{Joseph Konstan}, {and} \bibinfo{person}{John
  Riedl}.} \bibinfo{year}{2001}\natexlab{}.
\newblock \showarticletitle{Item-based collaborative filtering recommendation
  algorithms}. In \bibinfo{booktitle}{\emph{WWW}}. \bibinfo{pages}{285--295}.
\newblock


\bibitem[\protect\citeauthoryear{Schafer, Frankowski, Herlocker, and
  Sen}{Schafer et~al\mbox{.}}{2007}]%
        {schafer2007collaborative}
\bibfield{author}{\bibinfo{person}{J~Ben Schafer}, \bibinfo{person}{Dan
  Frankowski}, \bibinfo{person}{Jon Herlocker}, {and} \bibinfo{person}{Shilad
  Sen}.} \bibinfo{year}{2007}\natexlab{}.
\newblock \showarticletitle{Collaborative filtering recommender systems}.
\newblock In \bibinfo{booktitle}{\emph{The Adaptive Web}}.
  \bibinfo{pages}{291--324}.
\newblock


\bibitem[\protect\citeauthoryear{Shi, Yu, Da, Chen, and Zeng}{Shi
  et~al\mbox{.}}{2019}]%
        {shi2019virtual}
\bibfield{author}{\bibinfo{person}{Jing-Cheng Shi}, \bibinfo{person}{Yang Yu},
  \bibinfo{person}{Qing Da}, \bibinfo{person}{Shi-Yong Chen}, {and}
  \bibinfo{person}{An-Xiang Zeng}.} \bibinfo{year}{2019}\natexlab{}.
\newblock \showarticletitle{Virtual-taobao: Virtualizing real-world online
  retail environment for reinforcement learning}. In
  \bibinfo{booktitle}{\emph{AAAI}}. \bibinfo{pages}{4902--4909}.
\newblock


\bibitem[\protect\citeauthoryear{Srivastava, Hinton, Krizhevsky, Sutskever, and
  Salakhutdinov}{Srivastava et~al\mbox{.}}{2014}]%
        {srivastava2014dropout}
\bibfield{author}{\bibinfo{person}{Nitish Srivastava},
  \bibinfo{person}{Geoffrey Hinton}, \bibinfo{person}{Alex Krizhevsky},
  \bibinfo{person}{Ilya Sutskever}, {and} \bibinfo{person}{Ruslan
  Salakhutdinov}.} \bibinfo{year}{2014}\natexlab{}.
\newblock \showarticletitle{Dropout: a simple way to prevent neural networks
  from overfitting}.
\newblock \bibinfo{journal}{\emph{The Journal of Machine Learning Research}}
  \bibinfo{volume}{15}, \bibinfo{number}{1} (\bibinfo{year}{2014}),
  \bibinfo{pages}{1929--1958}.
\newblock


\bibitem[\protect\citeauthoryear{Thompson}{Thompson}{1933}]%
        {thompson1933likelihood}
\bibfield{author}{\bibinfo{person}{William~R Thompson}.}
  \bibinfo{year}{1933}\natexlab{}.
\newblock \showarticletitle{On the likelihood that one unknown probability
  exceeds another in view of the evidence of two samples}.
\newblock \bibinfo{journal}{\emph{Biometrika}} \bibinfo{volume}{25},
  \bibinfo{number}{3/4} (\bibinfo{year}{1933}), \bibinfo{pages}{285--294}.
\newblock


\bibitem[\protect\citeauthoryear{Wu, Chen, Yang, Wang, Tan, Zhang, Xu, and
  Gai}{Wu et~al\mbox{.}}{2018}]%
        {wu2018budget}
\bibfield{author}{\bibinfo{person}{Di Wu}, \bibinfo{person}{Xiujun Chen},
  \bibinfo{person}{Xun Yang}, \bibinfo{person}{Hao Wang}, \bibinfo{person}{Qing
  Tan}, \bibinfo{person}{Xiaoxun Zhang}, \bibinfo{person}{Jian Xu}, {and}
  \bibinfo{person}{Kun Gai}.} \bibinfo{year}{2018}\natexlab{}.
\newblock \showarticletitle{Budget constrained bidding by model-free
  reinforcement learning in display advertising}. In
  \bibinfo{booktitle}{\emph{CIKM}}. \bibinfo{pages}{1443--1451}.
\newblock


\bibitem[\protect\citeauthoryear{Yang, Li, Wang, Wu, Tan, Xu, and Gai}{Yang
  et~al\mbox{.}}{2019}]%
        {yang2019bid}
\bibfield{author}{\bibinfo{person}{Xun Yang}, \bibinfo{person}{Yasong Li},
  \bibinfo{person}{Hao Wang}, \bibinfo{person}{Di Wu}, \bibinfo{person}{Qing
  Tan}, \bibinfo{person}{Jian Xu}, {and} \bibinfo{person}{Kun Gai}.}
  \bibinfo{year}{2019}\natexlab{}.
\newblock \showarticletitle{Bid optimization by multivariable control in
  display advertising}. In \bibinfo{booktitle}{\emph{SIGKDD}}.
  \bibinfo{pages}{1966--1974}.
\newblock


\bibitem[\protect\citeauthoryear{Zhang, Ren, and Wang}{Zhang
  et~al\mbox{.}}{2016a}]%
        {zhang2016optimal}
\bibfield{author}{\bibinfo{person}{Weinan Zhang}, \bibinfo{person}{Kan Ren},
  {and} \bibinfo{person}{Jun Wang}.} \bibinfo{year}{2016}\natexlab{a}.
\newblock \showarticletitle{Optimal real-time bidding frameworks discussion}.
\newblock \bibinfo{journal}{\emph{arXiv preprint arXiv:1602.01007}}
  (\bibinfo{year}{2016}).
\newblock


\bibitem[\protect\citeauthoryear{Zhang, Rong, Wang, Zhu, and Wang}{Zhang
  et~al\mbox{.}}{2016b}]%
        {zhang2016feedback}
\bibfield{author}{\bibinfo{person}{Weinan Zhang}, \bibinfo{person}{Yifei Rong},
  \bibinfo{person}{Jun Wang}, \bibinfo{person}{Tianchi Zhu}, {and}
  \bibinfo{person}{Xiaofan Wang}.} \bibinfo{year}{2016}\natexlab{b}.
\newblock \showarticletitle{Feedback control of real-time display advertising}.
  In \bibinfo{booktitle}{\emph{WSDM}}. \bibinfo{pages}{407--416}.
\newblock


\bibitem[\protect\citeauthoryear{Zhang, Yuan, and Wang}{Zhang
  et~al\mbox{.}}{2014}]%
        {zhang2014optimal}
\bibfield{author}{\bibinfo{person}{Weinan Zhang}, \bibinfo{person}{Shuai Yuan},
  {and} \bibinfo{person}{Jun Wang}.} \bibinfo{year}{2014}\natexlab{}.
\newblock \showarticletitle{Optimal real-time bidding for display advertising}.
  In \bibinfo{booktitle}{\emph{SIGKDD}}. \bibinfo{pages}{1077--1086}.
\newblock


\bibitem[\protect\citeauthoryear{Zhang, Zhang, Gao, Yu, Yuan, and Liu}{Zhang
  et~al\mbox{.}}{2012}]%
        {zhang2012joint}
\bibfield{author}{\bibinfo{person}{Weinan Zhang}, \bibinfo{person}{Ying Zhang},
  \bibinfo{person}{Bin Gao}, \bibinfo{person}{Yong Yu},
  \bibinfo{person}{Xiaojie Yuan}, {and} \bibinfo{person}{Tie-Yan Liu}.}
  \bibinfo{year}{2012}\natexlab{}.
\newblock \showarticletitle{Joint optimization of bid and budget allocation in
  sponsored search}. In \bibinfo{booktitle}{\emph{SIGKDD}}.
  \bibinfo{pages}{1177--1185}.
\newblock


\bibitem[\protect\citeauthoryear{Zhou, Zhu, Song, Fan, Zhu, Ma, Yan, Jin, Li,
  and Gai}{Zhou et~al\mbox{.}}{2018}]%
        {zhou2018deep}
\bibfield{author}{\bibinfo{person}{Guorui Zhou}, \bibinfo{person}{Xiaoqiang
  Zhu}, \bibinfo{person}{Chenru Song}, \bibinfo{person}{Ying Fan},
  \bibinfo{person}{Han Zhu}, \bibinfo{person}{Xiao Ma},
  \bibinfo{person}{Yanghui Yan}, \bibinfo{person}{Junqi Jin},
  \bibinfo{person}{Han Li}, {and} \bibinfo{person}{Kun Gai}.}
  \bibinfo{year}{2018}\natexlab{}.
\newblock \showarticletitle{Deep interest network for click-through rate
  prediction}. In \bibinfo{booktitle}{\emph{SIGKDD}}.
  \bibinfo{pages}{1059--1068}.
\newblock


\bibitem[\protect\citeauthoryear{Zhu, Jin, Tan, Pan, Zeng, Li, and Gai}{Zhu
  et~al\mbox{.}}{2017}]%
        {zhu2017optimized}
\bibfield{author}{\bibinfo{person}{Han Zhu}, \bibinfo{person}{Junqi Jin},
  \bibinfo{person}{Chang Tan}, \bibinfo{person}{Fei Pan},
  \bibinfo{person}{Yifan Zeng}, \bibinfo{person}{Han Li}, {and}
  \bibinfo{person}{Kun Gai}.} \bibinfo{year}{2017}\natexlab{}.
\newblock \showarticletitle{Optimized cost per click in taobao display
  advertising}. In \bibinfo{booktitle}{\emph{SIGKDD}}.
  \bibinfo{pages}{2191--2200}.
\newblock


\bibitem[\protect\citeauthoryear{Zhu, Li, Zhang, Li, He, Li, and Gai}{Zhu
  et~al\mbox{.}}{2018}]%
        {zhu2018learning}
\bibfield{author}{\bibinfo{person}{Han Zhu}, \bibinfo{person}{Xiang Li},
  \bibinfo{person}{Pengye Zhang}, \bibinfo{person}{Guozheng Li},
  \bibinfo{person}{Jie He}, \bibinfo{person}{Han Li}, {and}
  \bibinfo{person}{Kun Gai}.} \bibinfo{year}{2018}\natexlab{}.
\newblock \showarticletitle{Learning tree-based deep model for recommender
  systems}. In \bibinfo{booktitle}{\emph{SIGKDD}}. \bibinfo{pages}{1079--1088}.
\newblock


\end{thebibliography}

\end{document}